\documentclass{elsart5}

\usepackage{graphicx}
\usepackage{amssymb}

\begin{document}

\begin{frontmatter}

\title{Evidence for Multiple Phase Transitions in
La$_{1-x}$Ca$_{x}$CoO$_{3}$}

\author[kyoto,koeln]{M. Kriener\corauthref{cor1}}
\ead{mkriener@scphys.kyoto-u.ac.jp}
%\ead[url]{ http://www.ss.scphys.kyoto-u.ac.jp/index.html.en}
\author[koeln]{M. Braden}
\author[koeln]{D. Senff}
\author[koeln]{O. Zabara}
\author[koeln]{T. Lorenz}

\address[kyoto]{Quantum Materials Lab, Department of Physics,
Graduate School of Science, Kyoto University}

\address[koeln]{University of Cologne, Z\"{u}lpicher Str. 77, 50937 K\"{o}ln, Germany}

\corauth[cor1]{}
\received{12 June 2005}
\revised{13 June 2005}
\accepted{14 June 2005}

\begin{abstract}
We report thermal-expansion and specific-heat data of  the series La$_{1-x}$Ca$_{x}$CoO$_{3}$ for $0\leq x \leq 0.3$. For $x=0$ the thermal-expansion coefficient $\alpha(T)$ features a pronounced maximum around $T=50$\,K caused by a temperature-dependent spin-state transition from a low-spin state ($S=0$) at low temperatures towards a higher spin state of the Co$^{3+}$ ions. The partial substitution of the La$^{3+}$ ions by divalent Ca$^{2+}$ ions causes drastic changes in the macroscopic properties of LaCoO$_{3}$. Around $x\approx 0.125$ the large maximum in $\alpha(T)$ has completely vanished. With further increasing $x$ three different anomalies develop.
\end{abstract}

\begin{keyword}
\PACS 75.40\sep 65.70 \sep 
\KEY  specific heat \sep thermal expansion \sep spin-state transition \sep metal-insulator transition \sep double-exchange mechanism
\end{keyword}

\end{frontmatter}

\section{Introduction}
LaCoO$_{3}$ crystallizes in a rhombohedral symmetry (R$\bar{3}$c) and has quite unusual physical properties. At low temperatures it features a nonmagnetic and insulating ground state. With increasing temperature a paramagnetic moment develops which causes a maximum in the susceptibility around 100\,K \cite{zobel02a,baier05a}. This is usually explained in terms of a temperature-driven spin-state transition of the Co$^{3+}$ ions which can occur in three different spin configurations: a low-spin (LS) ($t_{2g}^{6}e_{g}^{0}$, $S=0$), an intermediate-spin (IS) ($t_{2g}^{5}e_{g}^{1}$, $S=1$), or a high-spin state (HS) ($t_{2g}^{4}e_{g}^{2}$, $S=2$). At low temperatures a LS ground state is realized. The nature of the spin state at higher temperatures is controversially discussed, e.\,g. \cite{zobel02a,baier05a,itoh94a,senaris95a,korotin96a}. 

The spin-state transition causes also a large additional Schottky-like contribution to the thermal expansion of LaCoO$_{3}$ \cite{zobel02a}. The substitution of La$^{3+}$ by an amount $x$ of divalent Ca, Sr, or Ba ions influences this contribution and leads to a strong suppression of the spin-state transition. Furthermore all doping series show long-range ferromagnetic order and metallic behavior at higher doping levels $x \gtrsim 0.2$. In the case of Ca doping, ferromagnetic behavior is observed for the insulating compounds $x\leq 0.2$, too \cite{kriener04a}.

\section{Thermal Expansion and Specific Heat}
\begin{figure}[b]
\centering
\includegraphics[scale=0.775]{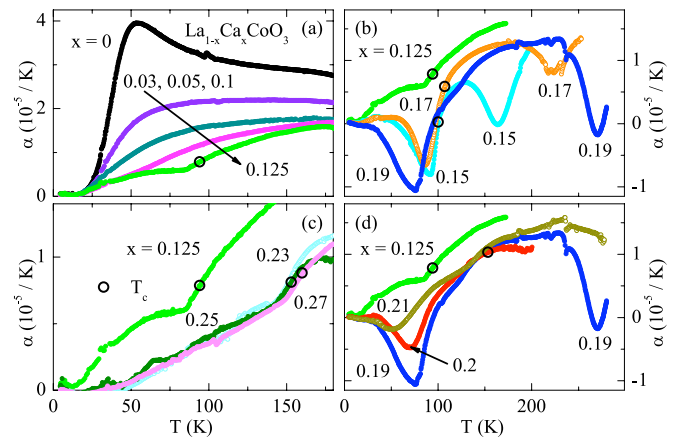}
\caption{Thermal expansion of La$_{1-x}$Ca$_{x}$CoO$_{3}$ for (a) $x \leq 0.125$, (b) $0.125 \leq x \leq 0.19$, (c) $0.23 \leq x \leq 0.27$ and (d) $0.19 \leq x \leq 0.21$. The curve for $x=0.125$ is shown in each panel. The black circles denote the critical temperatures $T_{c}$ of the onset of ferromagnetic order which have been taken from the magnetization data given in Ref.\ \cite{kriener04a}.}
\label{tad}
\end{figure}
In Fig.\ \ref{tad} the temperature dependence of the linear thermal-expansion coefficient $\alpha=1/L\cdot\partial L/\partial T$ of La$_{1-x}$Ca$_{x}$CoO$_{3}$ is shown. Fig.\  \ref{tad}\,(a) clearly reveals that the pronounced maximum in the thermal-expansion data of the undoped compound ($x=0$) is strongly suppressed with increasing doping concentration $x$. 
Above $x\gtrsim 0.125$ this indication of the spin-state transition has completely vanished. Instead several anomalies develop (Figs.\ \ref{tad}\,(b) -- (d)): The curves for $0.15 \leq x \leq 0.21$ exhibit two distinct minima and for $x \geq 0.19$ additional kinks occur. The low-temperature minimum (LTM) shifts with increasing $x$ to lower whereas the high-temperature minimum (HTM) moves to higher temperatures. For $x=0.21$ we could observe only the onset of the decrease in $\alpha(T)$, because the anomaly has shifted out of the available temperature interval, i.\,e.\ $T<300$\,K, for higher doping concentrations.
\begin{figure}[t]
\centering
\includegraphics[scale=0.63]{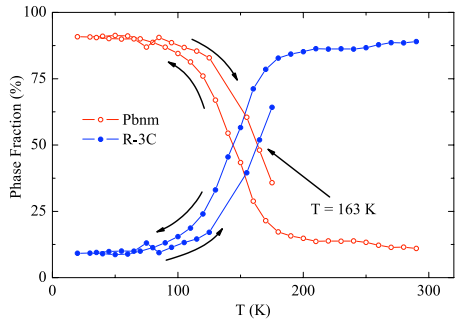}
\caption{Fraction of rhombohedral and orthorhombic phases as a function of $T$ obtained from X-ray diffraction data of La$_{0.85}$Ca$_{0.15}$CoO$_{3}$: The arrows indicate whether the data was taken with increasing or decreasing temperature. The lines are to guide the eyes.}
\label{roentg}
\end{figure}

At room temperature a doping induced structural phase transition from rhombohedral (R$\bar{3}$c, $x<0.2$) to orthorhombic symmetry (Pbnm, $x\geq 0.2$) is observed \cite{kriener04a}. This transition shifts to lower temperatures for smaller $x$. Fig.\ \ref{roentg} shows the temperature dependence of the structural phase fractions for $x=0.15$ which was obtained from X-ray diffraction data. It reveals a first-order phase transition. The HTM for $x=0.15$ occurs at $T_{min}^{\alpha}\approx 160$\,K which coincides with the crossover from R$\bar{3}$c to Pbnm in the X-ray data. Therefore the occurrence of the HTM in $\alpha(T)$ can be attributed to a doping-induced structural phase transition.
In Fig.\ \ref{tad}, the circles ($\circ$) mark the onset of ferromagnetic order. The critical temperatures $T_{c}$ of the magnetic transitions have been taken from the $M(T)$ data given in Ref.\ \cite{kriener04a}. They match the occurrence of the kinks in $\alpha(T)$ for $x=0.125$ and $x\geq 0.19$ (Figs.\ \ref{tad}\,(c), (d)). Therefore we conclude that the ''kink anomalies'' in the thermal-expansion data result from the onset of long-range ferromagnetic order.
\begin{figure}[t]
\centering
\includegraphics[scale=0.745]{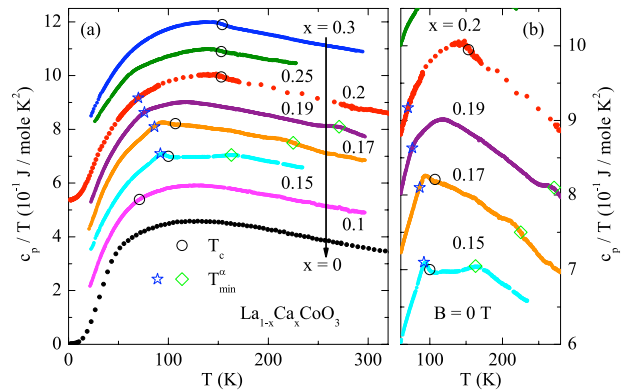}
\caption{Specific Heat $c_{p}/T$ of La$_{1-x}$Ca$_{x}$CoO$_{3}$: $\star$ and $\Diamond$ mark the occurrence of the LTM and HTM in $\alpha(T)$, $\circ$ signals the ferromagnetic phase transition. Panel (b) shows an enlarged view of the intermediate doping region $0.15\leq x \leq 0.2$. For clarity the different data sets are shifted by 0.1\,J/mol\,K$^{2}$ with respect to each other.}
\label{cpT}
\end{figure}

In Fig.\ \ref{cpT} specific-heat data $c_{p}/T$ of La$_{1-x}$Ca$_{x}$CoO$_{3}$ is presented. One can identify several anomalies, too, although they are much less pronounced than in $\alpha(T)$. The temperatures of the minima in $\alpha(T)$ (LTM: $\star$, HTM: $\Diamond$) and the onset of ferromagnetic order ($\circ$) are denoted. Fig.\ \ref{cpT}\,(b) gives an expanded view for $0.15\leq x \leq 0.2$. Please note, that for clarity the different data sets for $x>0$ are shifted by 0.1\,J/mol\,K$^{2}$ with respect to each other.

For $x\geq 0.15$ all compounds feature one or more weak anomalies in $c_{p}/T$, which coincide with the anomalies in $\alpha(T)$. For $x=0.15$ and 0.17 the LTM and the magnetic phase transition occur at similar temperatures. In the $\alpha(T)$ data of the higher doped specimens the onset of ferromagnetism is indicated by kinks. These kinks might be hidden by the large LTM anomalies for $x=0.15$ and 0.17. Therefore one cannot distinguish if the anomalies in $c_{p}/T$ for these two compounds are caused by the magnetic or a different transition. For $x= 0.19$ the specific-heat data features only a weak shoulder around the temperature of the corresponding LTM. The anomalies visible in the data of the higher doped compounds $x\geq 0.2$ are clearly related to the ferromagnetic phase transition. The structural phase transition shifts with increasing $x$ to higher temperatures and causes in both quantities clear anomalies. The appearance of the HTM in $\alpha(T)$ ($\Diamond$) and of the maxima at higher temperatures in $c_{p}/T$ take place at similar temperatures. For $x=0.2$ one can see a weak maximum in the specific-heat data above room temperature which agrees with the onset of a minimum in the thermal-expansion data. The origin of the LTM which is visible in $\alpha(T)$ is not clear yet. This anomaly vanishes for $x>0.21$ and possibly does not affect the specific heat. The investigation of its origin is still under progress and will be reported elsewhere.

In summary, we have studied the influence of hole doping on the series La$_{1-x}$Ca$_{x}$CoO$_{3}$ by thermal-expansion and specific-heat measurements. Depending on the concentration $x$ we observe up to three distinct anomalies. One of them is related to the ferromagnetic ordering, while a second one arises from a first-order structural phase transition. The origin of the third anomaly remains to be clarified.

\section{Acknowledgments}
This work was supported by the Deutsche Forschungsgemeinschaft through SFB 608.
%\bibliographystyle{/PaperBase/bst/efk_english}
%\bibliographystyle{/PaperBase/bst/elsart-num.bst}
%\bibliography{/PaperBase/preload,/PaperBase/ACoO3}

\end{document}